# Monitoring microplastics in live reef-building corals with microscopic laser particles

Vera M. Titze[1,2,†*], Jessica Reichert[3], Marcel Schubert[2], Malte C. Gather[1,2,4*]

[1] SUPA, School of Physics and Astronomy, University of St Andrews, Fife, Scotland
[2] Humboldt Centre for Nano- and Biophotonics, University of Cologne, Germany
[3] Hawaiʻi Institute of Marine Biology, University of Hawaiʻi at Mānoa, Hawaiʻi, Kāneʻohe, USA
[4] Cologne Excellence Cluster on Cellular Stress Responses in Aging-Associated Disease (CECAD), University of Cologne, Cologne, Germany
† Present address: Max Planck Institute of Colloids and Interfaces, Potsdam, Germany
*correspondence to: Vera.Titze@mpikg.mpg.de, Malte.Gather@uni-koeln.de

## Abstract

Micro- and nanoplastics pose a growing threat to marine organisms, such as reef-building corals. Yet, our understanding of microplastic uptake, interaction with coral tissue, and incorporation into coral skeletons remains limited, mainly due to the invasiveness of existing methods for detecting microplastics. Here, we exploit optical resonances in polymer spheres to transform microplastic particles into microscopic lasers. The bright, distinctive, and stable spectral signatures emitted by the microscopic laser particles function as optical barcodes, allowing extended tracking of microplastics transport through optically opaque coral tissue. Simultaneously, the lasers provide real-time sensing of dynamic changes at the microplastic surface with nanoscale resolution. Using confocal hyperspectral imaging, we establish the technical and analytical framework to capture coral anatomy and combine tracking and surface sensing into an integrated, non-invasive approach. With this, we explore the transport and internalization of individual microplastics in live corals, opening new avenues for understanding their ecological impact.

## Introduction

The emerging threat of plastic pollution in the world's oceans (*1*, *2*) has intensified efforts to understand its effects on marine life, particularly on increasingly endangered keystone species such as reef-building corals, which play a critical role in maintaining the structure and function of their ecological community. The impacts of plastic pollution are often species-specific (*3*, *4*) and can be exacerbated by additional stressors (*5*, *6*). Recent findings suggest that micro- and nanoplastics with dimensions < 20 µm, which are particularly challenging to detect and monitor, negatively affect the coral-symbiotic photosynthetic algae (*7*, *8*). Further, reef-building corals and other marine calcifiers may act as a long-term sink for micro- and nanoplastics by incorporating them into their calcium carbonate skeleton (*9*–*12*). However, the pathways of microplastic integration remain elusive, mainly due to limited understanding of their internalization and overgrowth. Previous studies have localized microplastics via invasive histological methods, which precludes *in vivo* monitoring of their precise spatial trajectories and thus their impact over time. Further challenges for *in vivo* monitoring arise from the optical opacity of the coral tissue and the wide range of spatial and temporal scales involved—from micrometre-sized particles and sub-second feeding responses to macroscopic coral colonies and calcification and overgrowth processes lasting weeks or months.

Here, we introduce a novel optical approach for tracking microplastics *in vivo* by using microplastic laser particles (M-LPs). We use spherical whispering gallery mode (WGM) microlasers composed of dye-doped polystyrene with a distribution in diameter centred around 16 µm to track the ingestion and trajectories of microplastics in living corals. Due to intrinsic optical resonances, M-LPs emit characteristic lasing spectra that can function as robust optical barcodes (*13*–*16*) and allow nanoscale sensing of the refractive index in their immediate environment (*17*–*20*). We find that M-LPs can act as a powerful model for microplastics, given their similar chemical composition and size. To follow trajectories of individual M-LPs, we integrate particle tracking with an advanced confocal hyperspectral imaging (CHSI) scheme, which provides information on the precise location of each particle within live coral tissue. In addition, by careful analysis of the spectrum of the emitted laser light, we can simultaneously monitor dynamic changes in the tissue surrounding the internalised microplastic particles. Studying the interaction of microplastics with coral tissue inherently must deal with macroscopic and optically dense samples. The exceptionally bright spectral signature of M-LPs provides resilience to signal degradation by absorption and scattering of light in coral tissue. Capitalising on this key advantage for deep-tissue measurements, we integrate particle tracking and *in vivo* sensing to reveal the internal pathways of microplastic transport in reef-building corals following particle uptake.

## Results

### Confocal hyperspectral imaging and spectral analysis

We first developed a protocol for capturing M-LP lasing spectra and extracting tracking and sensing information independently. To validate this protocol and demonstrate the principle of using M-LPs as a model for microplastics, we performed a "pulse exposure" experiment, in which we exposed live coral fragments to M-LPs (Fig. 1A, left). We then used CHSI to obtain 3D images of the coral anatomy and the M-LP positions while capturing the lasing spectra of each M-LP (Fig. 1A, right).

Next, we applied our analytical framework to the recorded lasing spectra, as is briefly explained here. The lasing spectra contain multiple peaks, and the positions of these peaks encode information on the diameter $d$ of the M-LP and on the refractive index $n_{ext}$ of the medium or tissue surrounding it (Fig. 1B,C). The dependence of the peak positions on $d$ is due to a spectral shift in WGM resonance with changing optical pathlength. The dependence on $n_{ext}$ is a consequence of the evanescent component of the electric field, which extends by several tens of nanometres from the resonator surface (Fig. 1B, bottom) and thus enables precise monitoring of the immediate environment of the M-LP. Importantly, the M-LPs support WGMs with two orthogonal polarisations, referred to transverse electric (TE) and transverse magnetic (TM), which differ in the spatial extent of their evanescent field. As a result, their sensitivity to changes in $n_{ext}$ differs, too, whereas both polarizations show a similar dependence on M-LP diameter. Consequently, we are able to determine whether measured shifts in peak positions originate from a change in $d$ or in $n_{ext}$, and fitting the peak positions in the lasing spectrum to a mathematical model in fact allows us to independently determine $d$ and $n_{ext}$ (Fig. 1D). Notably, this "spectral sizing" of M-LPs provides greatly enhanced precision in the measurement of $d$ (a few tens of nanometres or better) compared to existing *in vivo* imaging techniques, which are typically diffraction-limited. Combined with the natural size distribution of the M-LPs and the fact that the size of each M-LP remains constant over time, this high level of precision in size measurements enabled the use of the particle diameter as a unique identifier for each particle. By contrast, the refractive index in the surrounding the M-LP can change over time and our spectral analysis quantitatively tracks the evolution of $n_{ext}$. Finally, we linked the information on $d$ and $n_{ext}$ from the spectral analysis with positional information from the CHSI, which taken together enabled reliable particle tracking and refractive-index sensing, even in large, highly light-scattering coral samples.

### Localization and tracking of M-LPs *in vivo*

To simulate a microplastic exposure of corals, we exposed fragments of the common reef-building species *Porites lutea*, *Stylophora pistillata*, and *Acropora muricata* to an M-LP suspension for 24 h each (Methods). Immediately after the exposure, we performed CHSI to record the positioning of the M-LPs with respect to the coral epidermis (Fig. 2A), allowing us to investigate potential internalisation of microplastic. Simultaneously, we used spectral sizing of the M-LPs to monitor the presence and trajectories of the M-LPs (Fig. 2B). In all tested species, we found particles adhering to the coral surface (Fig. 2A-C). These M-LPs remained clearly identifiable by their lasing spectra even after several days (e.g., 11 days in *P. lutea*, Fig. 2D), demonstrating the excellent stability of the optical barcodes for long-term microplastic tracking. Over the time of the experiment, the corals effectively removed most M-LPs from healthy tissue, and particles that remained attached for longer were found only in areas with compromised tissue health. We found no evidence that the studied particles were internalized by corals directly during a pulse exposure.

**Refractive index sensing of immediate environment of M-LPs**

Next, we combined microplastic tracking with *in vivo* deep-tissue sensing to investigate feeding as a potential internalization pathway with significant microplastic uptake and retention. In this experiment, we introduced M-LPs using live *Artemia* nauplii as a delivery vector; for this, nauplii were first enriched with M-LPs and then fed to the heterotrophic coral species *S. pistillata* (Fig. 3A, Methods). We then monitored the coral sample for ~3 hours through continuous CHSI. Following *Artemia* ingestion by a subset of the polyps, M-LP signals were repeatedly detected during digestion and transport of the prey. The signal was at times intermittently lost due to optical scattering and absorption, especially when M-LPs were located at greater depth within the coral. Importantly, however, we were still able to reliably reidentify the different M-LPs via spectral sizing whenever the signal reemerged (Fig. 3B, Methods). Combining this information with multi-modal imaging data enabled reconstruction of trajectories of individual particles (Fig. 3C). For further analysis, the trajectories were then superimposed on an overview image of the entire fragment (Fig. 3C, trajectories labelled A to D). Confocal imaging confirmed that M-LPs were located beneath the heavily pigmented coral epidermis (Fig. 3D), where the active transport of the M-LPs by the coral became evident from the occurrence of motion artefacts (Fig. 3E, F).

We observed that particles ingested by a single polyp were transported to adjacent polyps and excreted there (Fig. 3C, e.g., M-LPs A and B), while other particles remained in the gastric cavity of the ingesting polyp before being egested (e.g., M-LP C). Furthermore, continuous tracking of M-LP A over an extended distance revealed pronounced blueshifts of the peaks in its lasing spectrum (Fig. 4A). Spectral modelling showed that these shifts originate from a gradual decrease in external refractive index—from $n_{ext}$ = 1.346 down to the refractive index of seawater ($n_{ext}$ = 1.338). This decline reflects a reduced concentration of high-refractive index materials such as proteins and lipids near the surface of the M-LP. Interestingly, the absolute values of the external refractive index were consistent across particles for the different stages of the digestion process: $n_{ext}$ = 1.345-1.350 shortly after ingestion, likely corresponding to partially digested *Artemia* in the gastric cavity; $n_{ext}$ = 1.342–1.345 within the gastrovascular canal system of the coenosarc tissue, where the organic material appears to get dispersed but the M-LP remains closely surrounded by coral tissue; and $n_{ext}$ = 1.338 during egestion of the M-LP from the gastric cavity (Fig. 4B). Refractive index changes during digestion were substantially larger in the feeding experiment ($\Delta n$ > 0.01) than in control measurements on undigested *Artemia* in seawater ($\Delta n$ < 0.002; Fig. 4C).

Linking the continuous decrease in refractive index during the feeding experiments to the positional tracking data suggests that the density of *Artemia* tissue in the immediate environment of the M-LP surface continues to decrease until the M-LP enters the gastric cavity of the neighbouring polyp (Fig. 4D). When the M-LP was egested from the gastric cavity, brightfield imaging of the particle and its surroundings again became feasible and confirmed that the coral had separated the M-LP from the *Artemia* tissue. Confocal imaging offered complementary insight into digestive processes, revealing localized increases in tissue motion during digestion and extended M-LP retention in bleached regions of the coral fragment (Fig. 4E).

## Discussion

We demonstrated that M-LPs are a multifunctional model for microplastics, providing intrinsic optical barcodes for tracking applications and acting as deep-tissue probes for sensing dynamic processes inside live corals. We show robust tracking of these microplastic particles by spectral sizing in scenarios where tracking by continuous imaging is not possible—specifically in multi-day experiments with intermediate imaging or in deep tissue where trajectories are partially obstructed. Such data enables precise localisation of microplastics or microplastic-contaminated food within the coral tissue and allows the identification and tracking individual particles across a range of timescales.

While we confirmed that corals can successfully remove microplastic particles from healthy tissue—in line with earlier reports (*4*)—our experiments show that the exposure to microplastics via food vectors such as *Artemia* leads to significant uptake even when microplastic is present at a substantially reduced average concentration. Following the uptake, we studied the transport of microplastic-contaminated food through the coenosarc tissue of corals through *in vivo* sensing of the digestive process. Here, the nanoscale spatial resolution of our sensing mechanism enabled us to selectively monitor processes on the M-LP surface, shedding light onto the dynamics of resource redistribution within a colony and the timescales of digestion. Notably, we find that uptake through contaminated food substantially increased the retention time of these particles. Our results highlight the importance of trophic transfer and bioaccumulation of microplastic particles in marine food webs, including mixotrophic organisms such as reef-building corals.

We expect future studies to extend measurement durations further, which could also benefit from interfacing M-LP detection with other emerging imaging modalities in the marine sciences (*21*–*23*). This work lays the foundation for exploring other M-LP materials, shapes and sizes, including alternative geometries such as nanodisks (*24*) and nanowires (*25*) and capitalising on recent progress in laser miniaturization (*26*). We also see great potential in exploring chemical functionalisation of microlasers (*19*) and nanoparticle-based biochemical sensing (*27*) to further advance microplastics research.

In summary, the combination of deep-tissue microplastic tracking with real-time measurements of microplastic-tissue interactions makes CHSI of M-LPs an attractive technique for investigating local responses of biological samples to the presence of microplastics. This dual functionality is highly relevant for ongoing efforts to explore microplastic interactions in organisms beyond reef-building corals (*28*), and our work provides the technical and analytical framework applicable to diverse biological contexts.

## Materials and Methods

### Multi-modal confocal imaging and collection of lasing spectra

Confocal images of corals, zooxanthellae autofluorescence, and M-LP fluorescence, as well as lasing spectra of M-LPs were collected with a confocal hyperspectral imaging (CHSI) setup that was built inhouse (*16*). The setup was based around a modified commercial confocal microscope (Nikon C1Si). It allowed to perform regular confocal fluorescence microscopy and, by using an alternative output light path, to acquire confocal hyperspectral images. Fluorescence microscopy used a 488 nm continuous wave (CW) laser delivering 0.15 mW incident power above the water surface of the sample dish and an inbuilt multi-channel detector of the confocal microscope with detection windows at 515 nm (bandwidth 30 nm, used for the GFP signal), 590 nm (bandwidth 50 nm, collecting fluorescence from M-LPs), and > 650 nm (650 nm long pass, suitable for zooxanthellae autofluorescence). For the confocal hyperspectral imaging, the M-LPs were excited with a 532 nm nanosecond pulsed laser (Coherent Helios, 125 kHz, 7.48 mW, 727 ps pulse duration). M-LP spectra were recorded by a line-scan camera (Teledyne Octoplus) connected to a grating spectrograph (Andor Shamrock SR500), which was fibre-coupled to the scan head of the microscope to collect high-resolution spectra ($\Delta\lambda$ = 66 pm) in the region from 586 nm to 614 nm. The system was further equipped with epifluorescence and transmission imaging, using a USB camera (Basler acA2040-55um) and also allowed direct sample observation through the eyepieces.

**Coral husbandry**

Fragments of the reef-building corals *Stylophora pistillata* (Esper, 1792), *Porites lutea* (Milne Edwards & Haime, 1851) and *Acropora muricata* (Linnaeus, 1758) (*n* = 1 per species) were kept in a 90 L seawater aquarium, together with other coral fragments (*Pocillopora verrucosa* (Ellis & Solander, 1786) and *Montipora spp*.), 2 damselfish (*Chromis viridis*), and ~ 30 small gastropods (*Euplica spp., Turbo spp*., and *Stomatella auricula*), which were obtained from the coral microcosm facility at Justus Liebig University Giessen, Germany. The animals were fed daily with ~1 $cm^3$ of frozen red plankton. The aquarium was illuminated with an LED lamp (Prime 16, Aquaillumination, 11 h:13 h light:dark cycle). A flow pump (Nero 3, Aquaillumination) was set to mimic a naturally varying water flow with random strengths between 8% and 54% of its maximum speed. The water was kept at a salinity of 35 ppt, a temperature of 26 °C, calcium content of 420 mg/L, and alkalinity of 7.5 dKH.

**Pulse exposure**

Pulse exposure measurements were performed on *Porites lutea*, *Stylophora pistillata*, and *Acropora muricata*. Coral fragments (*n* = 1 per species) were kept in individual 500 mL glass bottles to keep fragments in controlled conditions and at defined microplastic concentrations for ca. 24h. The bottles were filled with seawater from the main aquarium, supplemented with polystyrene microparticles (Microparticles GmbH, PS-FluoRed 15.5) to a final concentration of roughly $10^3$ particles per L, and kept in a water bath set to 26 °C. The same lamp used for the main aquarium was used to maintain similar light conditions. To minimize evaporation, each bottle was connected to a flow of humidified air via an air inlet at the bottom of each bottle, thereby also keeping M-LPs suspended in the water column. After particle exposure, fragments were removed from the bottle and placed into a petri dish, where they were fully submersed in fresh seawater for imaging. They were subsequently quarantined in another glass bottle with clean seawater to avoid contamination of the main aquarium with microplastic particles.

**Preparation of *Artemia* with M-LPs**

*Artemia* eggs (Dupla Marin, Germany) were added to 400 mL of seawater from the main tank and kept for approximately 48 hours at 26 °C under constant illumination and perspiration using an air pump connected to a glass pipette. Viable *Artemia* were collected using a plastic Pasteur pipette. *Artemia* were incubated with a suspension of M-LPs for approximately 15 minutes, at which point some *Artemia* had already ingested several M-LPs. A washing procedure was developed to gently separate the *Artemia* from the non-internalised M-LPs: Placing a light source at the top of the dish attracted the motile *Artemia*, whereas non-ingested M-LPs sank to the bottom of the dish. A concentrated *Artemia* suspension was collected from the illuminated area and subsequently diluted by fresh seawater. This washing procedure was repeated twice to ensure all non-internalised M-LPs were removed from the final coral feed, which was confirmed by visual inspection of transmission and epifluorescence images (e.g. Fig. 3A). The sample was then immediately placed into a -20°C freezer to prevent excretion of microplastics by the *Artemia* and stored until the feeding measurements were performed (typically, after 14 hours).

**Feeding measurements**

Feeding measurements were performed under the CHSI microscope. Fragments of the heterotrophic coral species *S. pistillata* were placed in a Petri dish, where they were fully submersed in fresh, M-LP-free seawater. Using a small Pasteur pipette, *Artemia* nauplii containing M-LPs were added to the water directly above the coral fragment, from where they slowly sank onto the fragment. Once the coral began to ingest the food in its vicinity, CHSI was started in the area surrounding a polyp that actively ingested *Artemia* nauplii.

**Mathematical analysis of lasing spectra for combined tracking and sensing**

The positions of four central consecutive lasing peaks were automatically extracted from each spectrum using Gaussian fits. These positions were passed to a MATLAB algorithm (*16*, *17*, *19*, *29*) that models the expected peak positions in the spectrum using an asymptotic expansion model of Mie scattering theory. By comparing the experimental peak positions to this model and finding the best fit, we determined the diameter of each M-LP as well as its external refractive index. Multiple iterations of the fitting routine were performed, starting with a coarse fit with open bounds for diameter and refractive index. The M-LP diameter obtained from this allowed to group the spectra into sets that are likely to originate from the same M-LP. A second fit with the diameter bounds limited to a 40 nm-wide window around the previously established diameter was then performed for each spectrum associated with the respective M-LP. All diameters obtained from this second series of fits were averaged and this average size value was used as the centre of a 20 pm-wide diameter window for the final iteration of the fit, which then yielded the most accurate refractive index information as it limited the error in the size parameter, assuming all fitted spectra originated from the same M-LP with a constant diameter. Fits giving a large residual error in M-LP peak positions (> 100 pm) were assumed to originate from a different M-LP and therefore excluded from the final result.

## Acknowledgments

The authors thank Patrick Schubert and Vanessa Tirpitz for their support during the aquarium setup and Michael Kühl, Cesar Pacherres, and Swathi Murthy for fruitful discussions.

**Funding:**
Leverhulme Trust (RPG-2017-231)
European Union’s Horizon 2020 Framework Programme (FP/2014-2020) / ERC grant agreement no. 640012 (ABLASE)
ERC grant agreement no. 101043047 (HYPERION)
EPSRC (EP/P030017/1)
Humboldt Foundation (Alexander von Humboldt professorship)
Instrument funding by the Deutsche Forschungsgemeinschaft in cooperation with the Ministerium für Kunst und Wissenschaft of North Rhine-Westphalia (INST 216/1120-1 FUGG)

**Author contributions:**

Conceptualization: VMT, JR, MS, MCG
Methodology: VMT, JR, MS, MCG
Investigation: VMT
Visualization: VMT, MS
Supervision: MS, MCG
Writing—original draft: VMT
Writing—review & editing: JR, MS, MCG

**Competing interests:** All other authors declare they have no competing interests.

## Figures and Tables

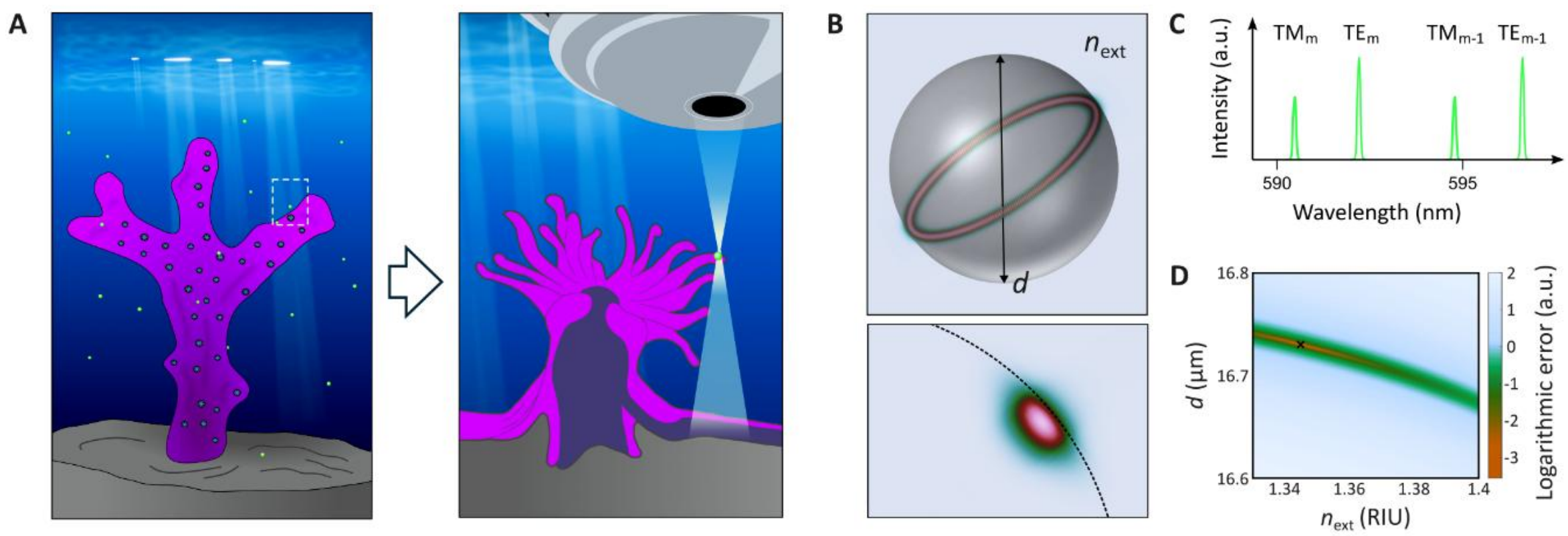

**Fig. 1. Generating optical barcodes from intrinsic optical resonances in M-LPs. (A**), Schematic of the experimental workflow: Corals are exposed to M-LPs, which serve as a model for microplastic particles (left). The CHSI system records 3D information on coral anatomy, along with the position and lasing spectra of M-LPs (right). (**B**), The electric field distribution of optical resonances inside an M-LP (top, simulation) depends on the particle diameter *d* and the refractive index of the surrounding medium $n_{ext}$. The sensitivity to $n_{ext}$ arises from the evanescent component of the electric field (bottom; dotted line indicates M-LP surface). (**C**), Schematic of the lasing spectrum of an M-LP, showing the characteristic multi-mode emission corresponding to optical resonances with different angular mode numbers *m* and polarizations (transverse electric, TE; transverse magnetic, TM). (**D**), Heatmap visualizing the residual error upon fitting a measured spectrum with a Mie scattering model to extract *d* and $n_{ext}$ (logarithmic colour scale). The global minimum is marked by a black cross.

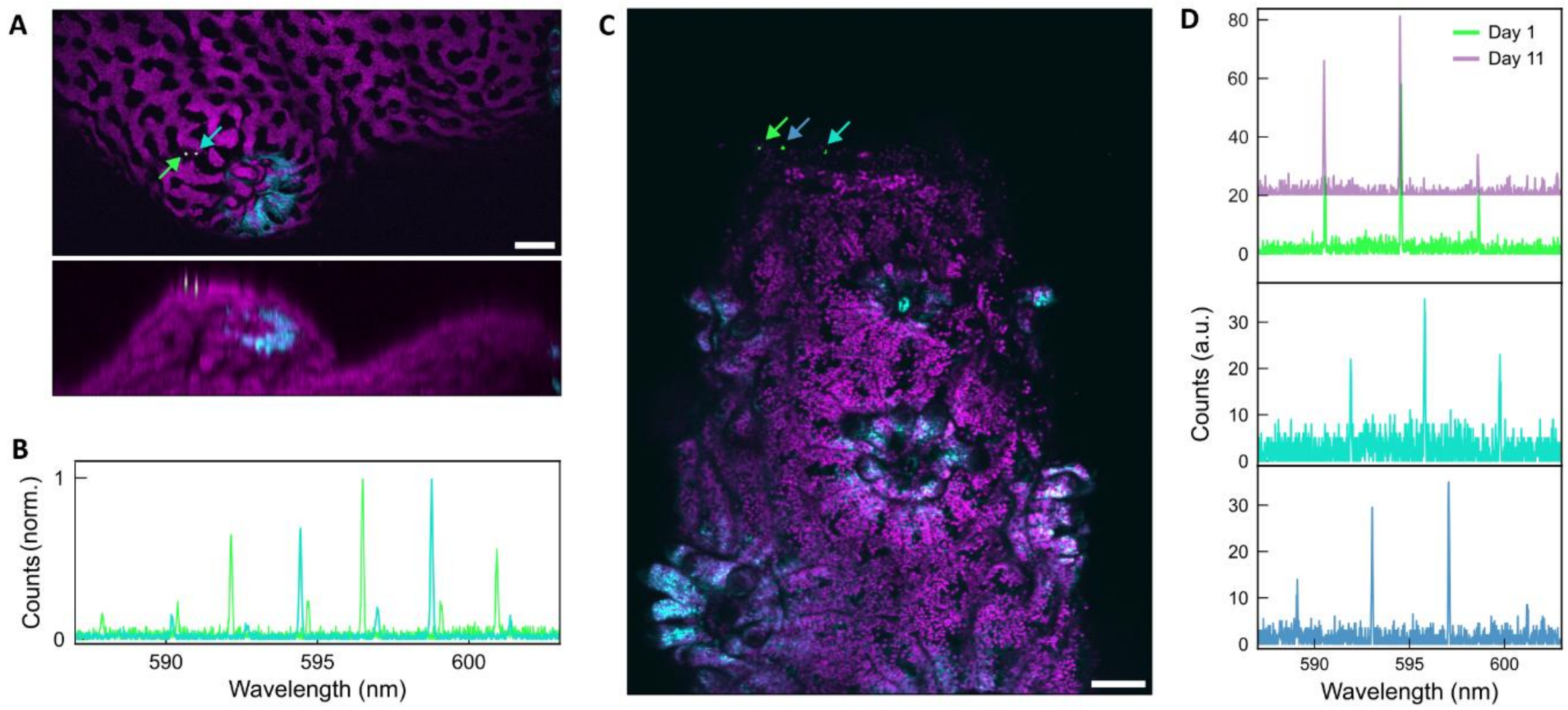

**Fig. 2. Optical barcodes made from M-LPs enable optical tracking of microplastic particles in live corals.** (**A**), Maximum intensity projection (MIP) of a CHSI volume stack of an A. muricata fragment in x-y (top) and x-z (bottom), showing zooxanthellae fluorescence (magenta), host autofluorescence (GFP, cyan) and two M-LPs attached to the coral epidermis (green, arrows). Scale bar, 200 µm. (**B**), Lasing spectra of M-LPs from a, illustrating the uniqueness of their spectra. (**C**), x-y-MIP of P. lutea (zooxanthellae, magenta; GFP, cyan) with three M-LPs (green) adhering to the tip of the fragment. Scale bar, 100 µm. (**D**), Spectra of the three M-LPs shown in c, recorded on first day after exposure (green traces). After 11 days, one particle was still attached and was re-identified by its spectrum (magenta trace, offset for clarity).

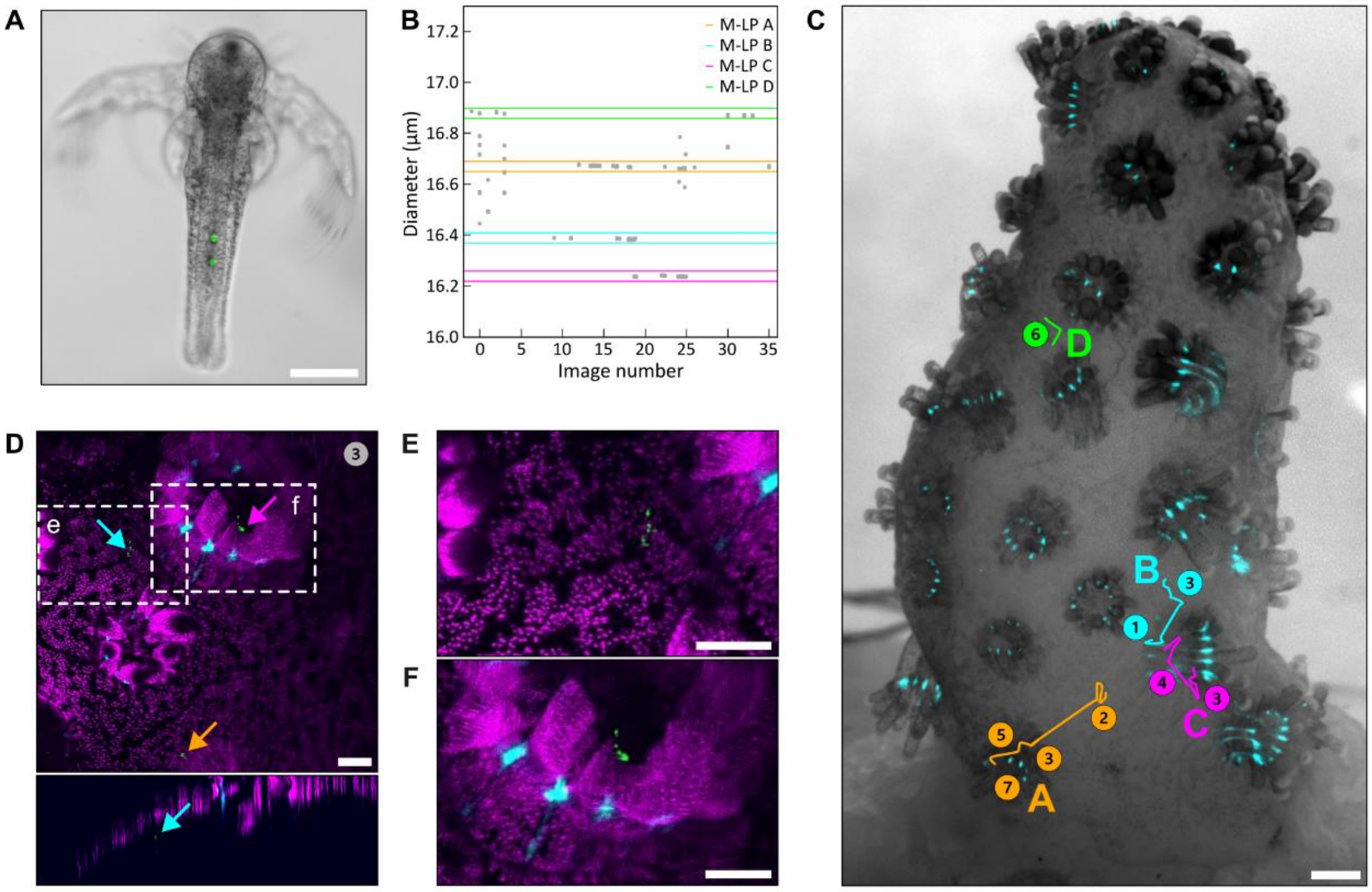

**Fig. 3. Investigating internal pathways of microplastics ingested during feeding.** (**A**), Combined brightfield transmission and epifluorescence image of *Artemia* (grey) containing M-LPs (green). Scale bar, 150 µm. (**B**)**,** Identifying M-LPs based on fitted diameters (grey dots) for all spectra from the entire measurement. Apparent clusters in particle diameters were enclosed by 40 nm-wide bands (coloured lines termed M-LP A to M-LP D) and associated spectra from each band were stored for subsequent analysis. (**C**), Brightfield stereomicroscope image of a coral fragment (grey) overlayed with GFP autofluorescence (cyan). Trajectories of the four tracked M-LPs from B are superimposed, with numbers indicating particle positions at defined time points. Scale bar, 500 µm. (**D**)**,** Representative confocal fluorescence image used for co-localization of coral anatomy and M-LP positions. x-y (top) and x-z (bottom) MIP, showing the zooxanthellae autofluorescence (magenta), host GFP autofluorescence (cyan), and the microplastic particles (green). M-LPs are marked with arrows, using the same colour-coding as in B and C. The time point is indicated by the number in the top right corner. Scale bar, 250 µm. (**E**-**F**), Magnified MIPs of areas around the M-LPs from D as indicated by corresponding boxes. The streaks of dots for M-LP B and C (cyan and magenta arrows in panel D) result from motion artefacts. Scale bars, 250 µm.

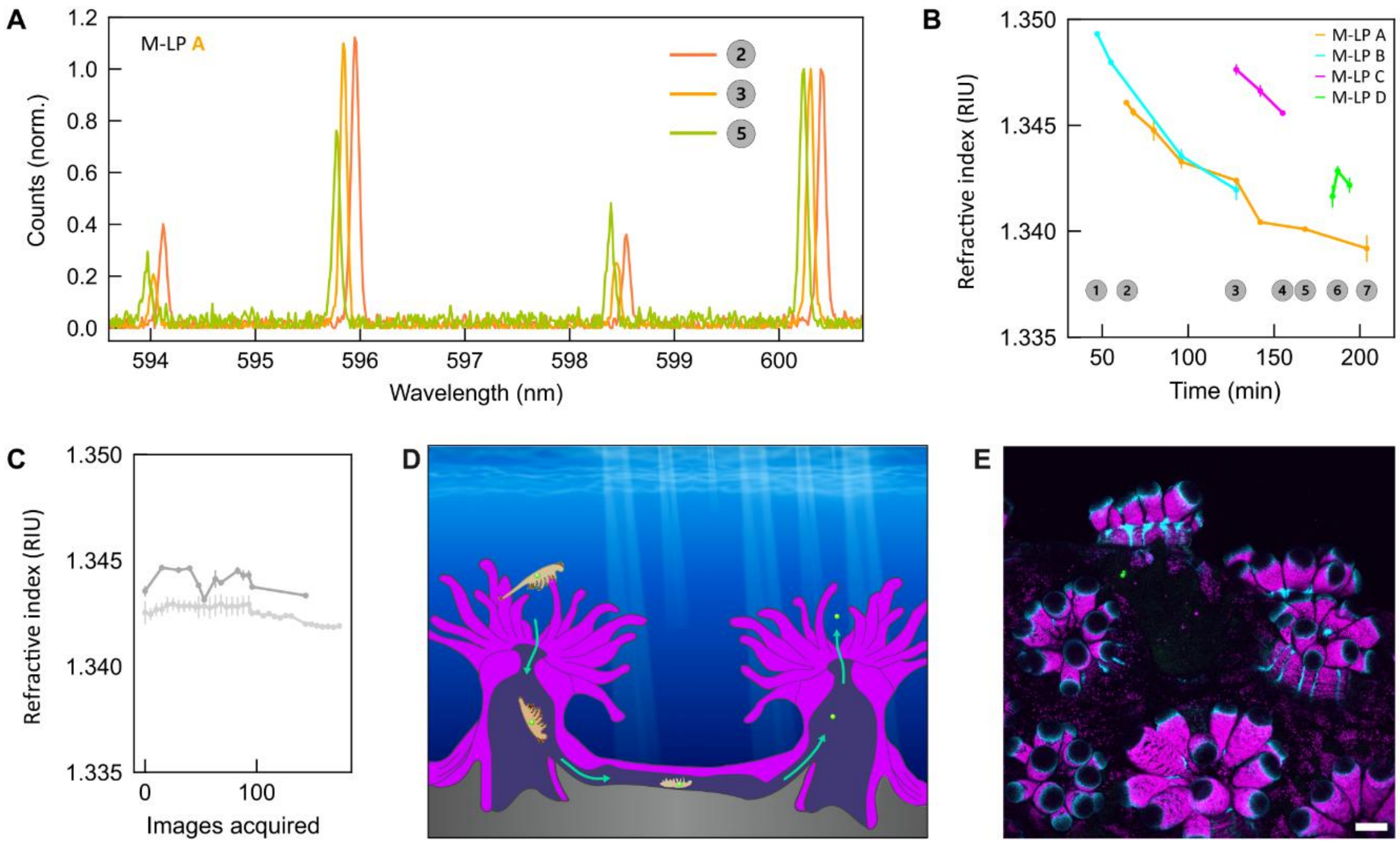

**Fig. 4. Digestion and transport dynamics revealed by the M-LP nanosensing mechanism.** (**A**), Lasing spectra from M-LP A at three different time points with numbers corresponding to c and d of Fig. 3, showing spectral shifts due to changes in external refractive index $n_{ext}$. (**B**), $n_{ext}$ obtained from analysis of lasing spectra for all tracked particles over the course of the experiment, showing distinct ranges that correspond to different stages of *Artemia* digestion and positions of the M-LPs within the coral. Numbers correspond to time points indicated in C and D of Fig. 3. (**C**), Control measurements for two M-LPs inside *Artemia* kept in seawater. Error bars indicate standard deviations across all lasing spectra evaluated at the respective time point. (**D**), Schematic of microplastic transport between two adjacent polyps through the coenosarc as reconstructed by analysis of the lasing spectra. *Artemia* nauplii containing M-LPs are fed to the coral, digested in the gastrovascular cavity, and M-LPs and *Artemia* derivatives get transported through the coenosarc. (**E**), Confocal x-y-MIP of a coral fragment fed with *Artemia* containing M-LPs, acquired on a commercial confocal microscope, showing the autofluorescence of GFP (cyan) and the zooxanthellae (magenta) as well as the M-LP emission (green). Scale bar, 250 μm.